# DeepMTS: Deep Multi-task Learning for Survival Prediction in Patients with Advanced Nasopharyngeal Carcinoma using Pretreatment PET/CT

Mingyuan Meng, Bingxin Gu, Lei Bi, *Member, IEEE*, Shaoli Song, David Dagan Feng, *Fellow, IEEE*, and Jinman Kim, *Member, IEEE*

*Abstract* — Nasopharyngeal Carcinoma (NPC) is a malignant epithelial cancer arising from the nasopharynx. Survival prediction is a major concern for NPC patients, as it provides early prognostic information to plan treatments. Recently, deep survival models based on deep learning have demonstrated the potential to outperform traditional radiomics-based survival prediction models. Deep survival models usually use image patches covering the whole target regions (e.g., nasopharynx for NPC) or containing only segmented tumor regions as the input. However, the models using the whole target regions will also include non-relevant background information, while the models using segmented tumor regions will disregard potentially prognostic information existing out of primary tumors (e.g., local lymph node metastasis and adjacent tissue invasion). In this study, we propose a 3D end-to-end Deep Multi-Task Survival model (DeepMTS) for joint survival prediction and tumor segmentation in advanced NPC from pretreatment PET/CT. Our novelty is the introduction of a hard-sharing segmentation backbone to guide the extraction of local features related to the primary tumors, which reduces the interference from non-relevant background information. In addition, we also introduce a cascaded survival network to capture the prognostic information existing out of primary tumors and further leverage the global tumor information (e.g., tumor size, shape, and locations) derived from the segmentation backbone. Our experiments with two clinical datasets demonstrate that our DeepMTS can consistently outperform traditional radiomics-based survival prediction models and existing deep survival models.

## I. INTRODUCTION

Nasopharyngeal Carcinoma (NPC) is a malignant epithelial cancer arising from the nasopharynx, which is the upper part of the pharynx and connects with the nasal cavity above the soft palate [1]. According to the GLOBOCAN 2020 published by the International Agency for Research on Cancer (IARC), there are approximately 133,354 new NPC patients and 80,008 NPC-related deaths worldwide in 2020 [2]. Survival prediction, a regression task that models the survival outcomes of patients, is a major concern for NPC patients, as it provides early prognostic information that is needed to guide treatments. In clinical practice, survival prediction is especially essential for patients with advanced NPC because the widely used TNM staging system has limited prognostic value for advanced NPC [3]. Consequently, the current research trend has focused on advanced NPC [8][18]. Survival prediction is an intractable challenge as it has to take into account incomplete survival data. Generally, survival data includes many right-censored samples, for which the exact time of events occurring is unclear. For example, patients may be lost to follow-up, and it is only known that the events did not occur in a period of time. Survival prediction models, to make the maximum use of existing information, need to learn from both complete (uncensored) and incomplete (censored) samples.

Radiomics, as a widely recognized computational method for survival prediction, refers to the extraction and analysis of high-dimensional quantitative features from medical images [6]. It has been widely used for survival prediction in various cancers including NPC [4][7][8]. Radiomics is heavily dependent on human intervention and prior knowledge, such as the need for manual segmentation, handcrafted feature extraction, and manual tuning of survival models. Therefore, its limitation in bringing a source of human bias has been well recognized [9][10]. There have been attempts at automating some steps in the radiomics pipeline to reduce human bias, such as automating the segmentation step [11][12]. Nevertheless, radiomics still has another limitation: radiomics features are extracted from segmented tumor regions, which are usually limited to primary tumor regions [4][8]. However, some potentially prognostic information, such as local lymph node metastasis and adjacent tissue invasion, exists in the surrounding regions out of primary tumors [13]. This means radiomics features cannot represent the

Manuscript received October 31, 2021; revised May 01, 2022; accepted June 02, 2022. This work was supported in part by Australian Research Council (ARC) grants (DP200103748 and IC170100022).

M. Meng, L. Bi, D. Feng, and J. Kim are with the School of Computer Science, the University of Sydney, Sydney, Australia.
B. Gu and S. Song are with the Department of Nuclear Medicine, Fudan University Shanghai Cancer Center; Department of Oncology, Shanghai Medical College, Fudan University; Center for Biomedical Imaging, Fudan University; Shanghai Engineering Research Center of Molecular Imaging Probes; Key Laboratory of Nuclear Physics and Ion-beam Application (MOE), Fudan University; Shanghai, China.
Corresponding to L. Bi and J. Kim. (email: lei.bi@sydney.edu.au and jinman.kim@sydney.edu.au)
M. Meng and B. Gu contributed equally to this work.

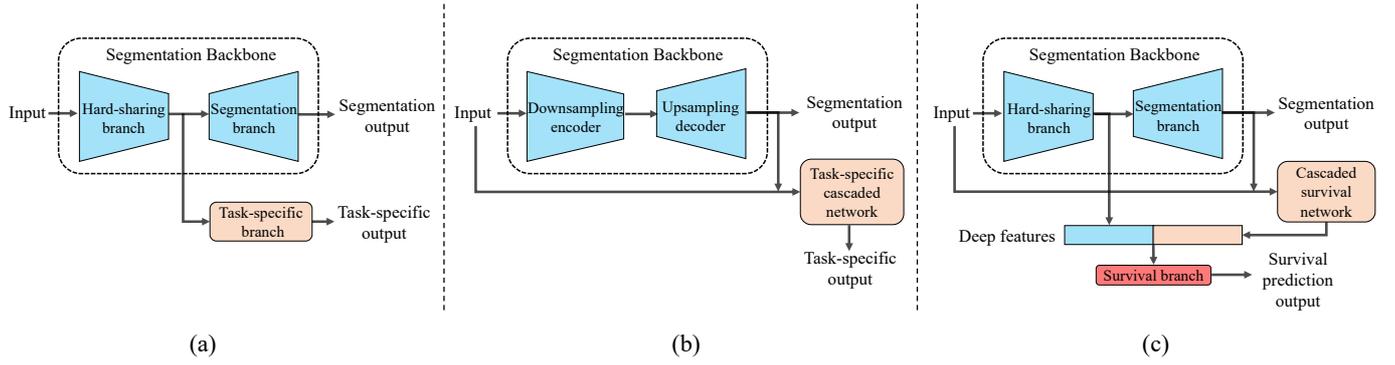

Fig. 1. A conceptual illustration of (a) the hard-sharing multi-task architecture, (b) the cascaded multi-task architecture, and (c) our hybrid multi-task architecture used in the DeepMTS for joint survival prediction and tumor segmentation.

prognostic information existing out of primary tumors and also lose global anatomical information such as tumor location. This limitation is even more severe for advanced NPC, because many vital tissues and organs adjacent to the nasopharynx (e.g., brain, ethmoidal sinus, and orbit) might have already been invaded by advanced NPC [14]. There have been attempts at enabling radiomics features to represent the information in tumor surroundings, such as using shell features extracted from outer voxels around the tumor boundary [15]. However, these attempts are limited to local tumor surroundings and still cannot represent global anatomical information.

Deep learning, which leverages Deep Neural Networks (DNNs) to learn representations of image patterns, has been used for survival prediction in various cancers including NPC [10][18][19]. Deep survival models using DNNs can directly predict survival outcomes from image input, in which the feature extraction and feature analysis are jointly learned in an end-to-end manner [19][23][24]. End-to-end deep survival models allow for automatic learning of relevant features without human intervention. Therefore, they can remove the human bias introduced by handcrafted features and potentially discover high-level semantic features that may be overlooked by manually-defined feature extraction [25]. Furthermore, end-to-end deep survival models usually do not need segmentation of primary tumors and take as input image patches covering the whole target regions (e.g., nasopharynx for NPC) [19][24], which enables them to use the prognostic information existing in the whole target regions. However, existing deep survival models do not make any assumptions or impose any constraints on tumor regions [19][24]. Therefore, non-relevant background information is unavoidably fed into the models and interferes with the prediction process. There are deep survival models that attempt to take the segmented primary tumor regions as input [23], which reduces the interference from non-relevant background information. Unfortunately, these models disregard the potentially prognostic information out of primary tumors.

Multi-task learning, in which multiple related tasks are simultaneously learned by a shared model [26][27], has the potential to overcome the above-mentioned limitations. Recently, tumor segmentation has been used as an auxiliary task in deep multi-task learning and benefited tumor-related clinical tasks such as genotype prediction [29], treatment response prediction [30], and clinical score prediction [35]. In most cases [30][34][35], a hard-sharing multi-task architecture was used: a segmentation network (U-Net [32] or its many variants) is used as a backbone, in which the downsampling encoder is used as a hard-sharing branch to extract common features shared by all tasks and then is followed by multiple task-specific output branches (Fig. 1a). With the hard-sharing architecture, the tumor segmentation task can implicitly guide the model to extract local features related to primary tumor regions [30], which allows the model to focus on primary tumors and get less interference from non-relevant background information. Moreover, a cascaded multi-task architecture has also been used [29]: the output of the segmentation backbone and the original input are fed into a task-specific cascaded network, and the two networks are trained cooperatively in an end-to-end manner (Fig. 1b). With the cascaded architecture, the output of the segmentation backbone, providing global tumor information (e.g., tumor size, shape, and locations), is explicitly leveraged by the task-specific cascaded network to enhance performance.

In this study, we propose a deep multi-task framework for joint tumor segmentation and survival prediction. Our novelty is to incorporate tumor segmentation, as an auxiliary task, for survival prediction using a hybrid multi-task architecture. The hybrid multi-task architecture extends the commonly used hard-sharing multi-task architecture (Fig. 1a) and cascaded multi-task architecture (Fig. 1b). Specifically, in our hybrid multi-task architecture, deep features are derived from both a hard-sharing segmentation backbone and a cascaded survival network, and we then feed the deep features into a survival branch for survival prediction (Fig. 1c). Our hybrid multi-task architecture can benefit from the advantages of these two baseline architectures and make use of tumor segmentation information both implicitly (through the hard-sharing part) and explicitly (through the cascaded part). The hard-sharing backbone can extract local features related to the primary tumor regions, while the cascaded survival network can capture the prognostic information existing out of primary tumors (from the original input) and further leverage the output of the hard-sharing backbone to derive global tumor information.

Under the proposed framework, we designed a customized segmentation network as the backbone and established a 3D end-to-end Deep Multi-Task Survival model (DeepMTS). Our DeepMTS simultaneously predicts disease progression risk and segments primary tumor regions using pretreatment PET/CT of advanced NPC patients. Manual segmentation of primary tumors was only used as labels for training and was not required for inferring unseen testing samples. The prediction of disease progression risk is a regression task, where a higher risk value indicates that the patient has shorter Progression-Free Survival (PFS). For clinical practice, the predicted risk can be used to stratify patients into different risk groups [19] or be used as a high-level prognostic factor to establish prognostic nomograms [8][18], which facilitates personalized treatment plans.

## II. RELATED WORK

### A. Traditional Survival Models

The Cox Proportional Hazards (CPH) model [37] is the most widely used survival model in medical applications. So far, the CPH model has been extended to many variants, such as Lasso-Cox [38] and EN-Cox [39] that use Least Absolute Shrinkage and Selection Operator (LASSO) regression and elastic-net penalizations for feature selection, respectively. Recently developed survival models include Random Survival Forest (RSF) [21] and Boosting Concordance Index (BoostCI) [28]. RSF is a tree model that generates an ensemble estimate for survival data. BoostCI is a survival model in which the concordance index metric is seen as an equivalent smoothed criterion using the sigmoid function. These traditional survival models cannot directly apply to image data, so image-derived handcrafted radiomics features are necessary for the traditional survival analysis pipeline.

### B. Deep Learning for Survival Prediction

Early studies of deep learning for survival prediction focus on learning the nonlinear relationships between clinical prognostic indicators and survival outcomes. Katzman et al. [20] proposed a deep survival model (DeepSurv) with a Cox negative logarithm partial likelihood loss. Gensheimer et al. [22] proposed a discrete-time survival model to directly predict the conditional probability of patients surviving on each time interval. Same as traditional survival models, these early deep survival models cannot directly apply to image data.

Recently, Zhang et al. [23] proposed an end-to-end CNN-based survival model (CNN-Survival) for survival prediction in pancreatic ductal adenocarcinoma using CT. CNN-Survival takes manually segmented primary tumor regions as input and thus disregards the prognostic information existing out of primary tumors. Besides, CNN-Survival is a 2D model that makes predictions based on only one single slice, thereby disregarding the potentially prognostic information that exists in 3D tumor volumes as well. Kim et al. [24] proposed a 3D end-to-end Deep Learning survival Prediction Model (DLPM) for lung adenocarcinomas. DLPM takes as input 3D patches covering the whole target regions (i.e., lung nodules and surrounding regions). However, as we have mentioned, non-relevant background information potentially interferes with the prediction process because DLPM did not make any assumptions or impose any constraints on the tumor regions.

### C. Survival Prediction in Patients with NPC

Image-derived survival prediction has been explored for NPC [4][8]. Zhang et al. [8] explored the survival prediction in NPC using MRI, in which radiomics features were extracted and then were used to build a Lasso-Cox model. Similarly, Lv et al. [4] also explored this problem using PET/CT, in which the most prognostic radiomics features were selected by univariate Cox analysis and then were used to build a CPH model.

Recently, deep learning has also been used for survival prediction in patients with NPC [18][19]. Peng et al. [18] proposed an early study where deep learning was introduced into the survival prediction in NPC patients. They used a pre-trained 2D CNN to extract deep features from PET and CT separately and then fed the deep features and radiomics features into a Lasso-Cox model to establish a prognostic nomogram. Their study suggested that deep features can serve as reliable and powerful indicators for survival prediction in NPC patients. Subsequently, Jing et al. [19] proposed the first 3D end-to-end deep model for survival prediction in NPC patients. It is a Multi-modality Deep Survival Network (MDSN) that directly predicts the risk of disease progression from pretreatment MRI. Their study demonstrated that end-to-end deep survival models are more effective to extract relevant features and showed higher prognostic performance. However, MDSN shares the same limitation as DLPM [24] because it takes as input the whole target regions (i.e., nasopharynx) and did not impose any constraints on its learning focus as well. Moreover, MDSN was designed for an MRI dataset of 1417 NPC patients. From our review, we have not identified any study where end-to-end deep survival models were used for PET/CT data of NPC patients.

### D. Deep Multi-task Learning for Medical Image Analysis

Deep multi-task learning has been recently explored for medical image analysis. Liu et al. [29] proposed a cascaded CNN for joint segmentation and genotype prediction of brainstem gliomas using MRI as well. Jin et al. [30] proposed a deep multi-task model to perform tumor segmentation and treatment response prediction using pre-/post-treatment MRI. Amyar et al. [34] proposed a deep multi-task model to jointly identify COVID-19 patients and segment lesions from chest CT. Cao et al. [35] proposed a multi-task neural network for joint hippocampus segmentation and regression of mini-mental state examination score using MRI. Most studies adopted the hard-sharing multi-task architecture [30][34][35], while others used the cascaded multi-task architecture [29].

For survival prediction, Andrearczyk et al. [31] recently proposed a deep multi-task U-Net for joint tumor segmentation and survival prediction in head and neck cancer. They also adopted the hard-sharing multi-task architecture which we identified as a baseline and hence introduce a hybrid multi-task architecture in this study. Andrearczyk et al. [31] demonstrated that the tumor segmentation task can facilitate the survival prediction task using deep multi-task learning.

## III. METHOD

### A. Overview

We propose DeepMTS to directly predict the risk of disease progression from PET/CT in an end-to-end manner. Fig. 2 shows the workflow of the proposed DeepMTS. The whole workflow takes a pair of preprocessed 3D PET/CT images ($128 \times 128 \times 112$; see Section IV-B) as input and predicts a disease progression risk and a tumor segmentation mask. The DeepMTS is composed of a segmentation backbone and a cascaded survival network (CSN). The segmentation backbone is a customized segmentation network based on 3D U-net [32] (in Section III-B) and the CSN is a modified 3D DenseNet [17] (in Section III-C). Deep features are derived from these two components and then are fed into several Fully-Connected (FC) layers for survival prediction. The whole architecture was trained in an end-to-end manner to minimize a combined loss including a segmentation loss $L_{seg}$ and a survival prediction loss $L_{sur}$ (in Section III-D).

Specifically, the preprocessed PET and CT images are first concatenated and fed into the segmentation backbone. Then, the segmentation backbone produces a tumor probability map where each voxel value represents the probability to be within the tumor regions. Finally, the tumor probability map and the preprocessed PET/CT images are concatenated and fed into the CSN. For the tumor segmentation task, the tumor probability map can be thresholded at 0.5 to obtain a binary tumor segmentation mask. For the survival prediction task, 124 deep features are derived from the segmentation backbone and fed into a Rectified Linear Unit (ReLU)-activated FC layer with 64 neurons (FC1); 112 deep features are derived from the CSN and also fed into a ReLU-activated FC layer with 64 neurons (FC2). The FC1, FC2, and clinical features $C$ (e.g., TNM stage, age, gender, etc.) are concatenated and fed into a non-activated (linear) FC layer with a single neuron (FC3). The output of FC3 is the predicted risk of disease progression. Moreover, dropout with 0.5 probability and L2 regularization with 0.1 coefficient are used in all FC layers.

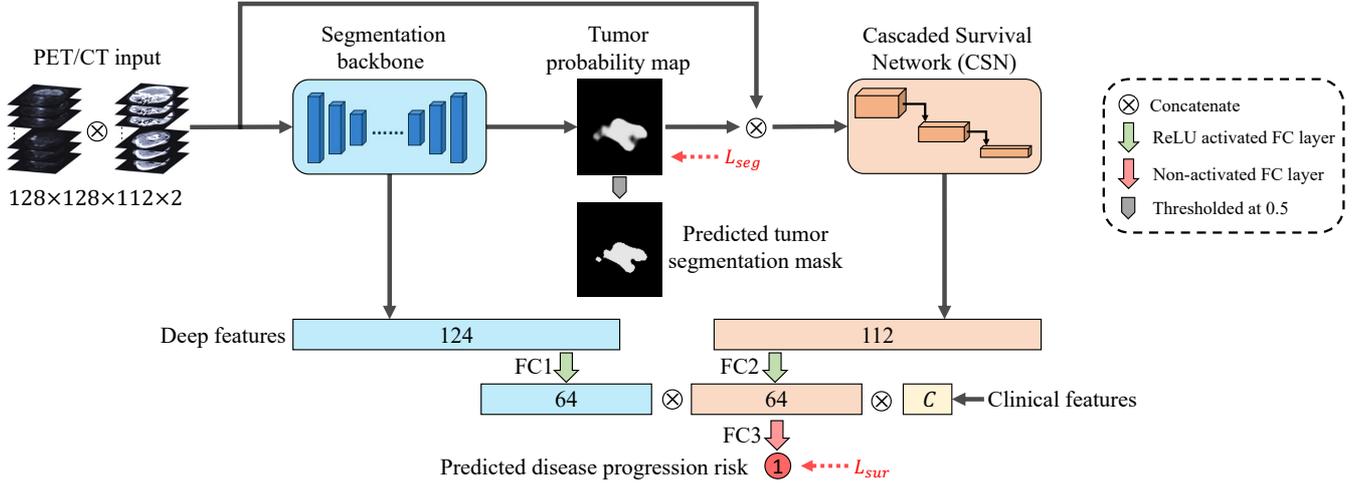

Fig. 2. The workflow of DeepMTS for joint survival prediction and tumor segmentation. It is composed of a segmentation backbone (blue rectangle) and a cascaded survival network (orange rectangle). Deep features are derived from these two components and fed into FC layers for survival prediction. Clinical features could be potentially prognostic clinical indicators such as age, gender, and TNM stage.

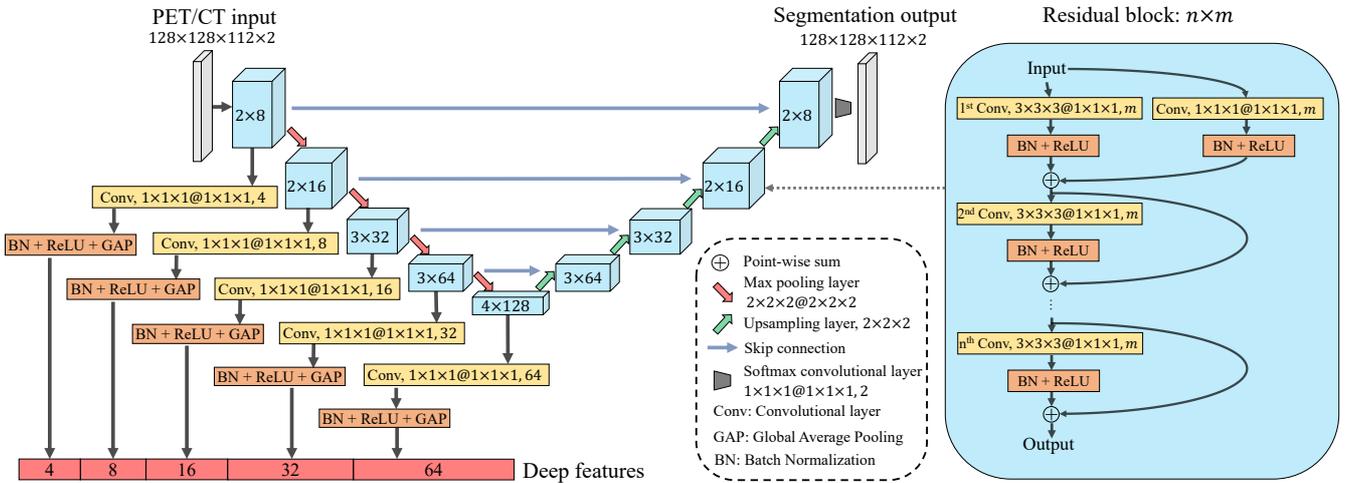

Fig. 3. The architecture of our segmentation backbone. It is composed of a downsampling hard-sharing branch and an upsampling segmentation branch. The residual block (blue cube) denoted with "$n \times m$" consists of $n$ stacked convolutional layers with a kernel number of $m$. The notation "$a \times b \times c @ d \times e \times f, m$" denotes a kernel size of $a \times b \times c$, a stride size of $d \times e \times f$, and a kernel number of $m$.

## B. Segmentation Backbone

The architecture of our segmentation backbone is illustrated in Fig. 3. It is a customized segmentation network based on 3D U-net [32] and consists of a downsampling hard-sharing branch and an upsampling segmentation branch. As shown in Fig. 3, the left half is the hard-sharing branch that extracts common features for both segmentation and survival prediction tasks; the right half is the segmentation branch that receives the common features through skip connections and performs tumor segmentation. Both two branches are composed of residual blocks (blue cubes in Fig. 3). Max pooling layers or upsampling layers with a kernel size of $2 \times 2 \times 2$ are used between adjacent residual blocks for downsampling or upsampling.

A residual block denoted by "$n \times m$" consists of $n$ stacked convolutional layers with a kernel size of $3 \times 3 \times 3$ and a kernel number of $m$. In this residual block, the 1st convolutional layer comes with a non-linear residual connection implemented by a convolutional layer with a kernel size of $1 \times 1 \times 1$ and a kernel number of $m$, while the rest $n-1$ convolutional layers come with shortcut residual connections. All convolutional layers are followed by Batch Normalization (BN) and ReLU activation. The last residual block is followed by a softmax activated convolutional layer with a kernel size of $1 \times 1 \times 1$ and a kernel number of 2. The output of the last convolutional layer is the predicted tumor probability map. Deep features are derived from multi-scale residual blocks in the hard-sharing branch for survival prediction. The output of each residual block is fed into a convolutional layer with a kernel size of $1 \times 1 \times 1$, followed by BN, ReLU activation, and a Global Average Pooling (GAP) layer. Finally, a total of 124 deep features are derived from 5 residual blocks.

## C. Cascaded Survival Network

The CSN is a modified version of 3D DenseNet [17], which is illustrated in Fig. S1 of the Supplementary Materials. Firstly, the preprocessed PET/CT and the tumor probability map (from the segmentation backbone) are concatenated and fed into a convolutional layer and a max-pooling layer for downsampling. Then, the feature maps are fed into 3 dense blocks and 3 transition blocks. The 3 dense blocks have 4, 8, and 16 bottleneck blocks, respectively. The bottleneck blocks in the same dense block are densely connected (refer to [17] for more details). Each bottleneck block consists of two convolutional layers with a kernel size of $1 \times 1 \times 1$ and $3 \times 3 \times 3$, respectively. BN and ReLU activation are used before and dropout with 0.05 probability is used after the convolutional layers in the bottleneck blocks. Each dense block is followed by a transition block consisting of BN, ReLU activation, a convolutional layer with a kernel size of $1 \times 1 \times 1$, and dropout with 0.05 probability. The first two transition blocks are followed by an average pooling layer with a kernel size of $2 \times 2 \times 2$ for downsampling and connecting with later dense blocks. Deep features are derived from the transition blocks for survival prediction. The output of each transition block is fed into BN, ReLU activation, and a GAP layer. Finally, a total of 112 deep features are derived from 3 transition blocks.

## D. Optimization

Our DeepMTS was trained in an end-to-end manner to minimize a combined loss $L$ as follows:
$$L = L_{seg} + L_{sur} + \lambda L_{reg}, \quad (1)$$
where $L_{seg}$ (Eq.2) is a loss function for the tumor segmentation task, $L_{sur}$ (Eq.3) is a loss function for the survival prediction task, and $L_{reg}$ with the coefficient $\lambda$ is the L2 regularization term used in FC layers.

For the tumor segmentation task, the $L_{seg}$ is a Dice loss [33]:
$$L_{seg} = -\frac{2\sum_i^N p_i g_i}{\sum_i^N p_i^2 + \sum_i^N g_i^2}, \quad (2)$$
where $p_i \in [0,1]$ is voxels of the predicted tumor probability map, $g_i \in \{0,1\}$ is the voxels of the ground truth tumor segmentation mask (label), and the sums run over all $N$ voxels of segmentation space.

For the survival prediction task, we used a Cox negative logarithm partial likelihood loss [20] as the $L_{sur}$ to handle right-censored survival data, which is shown as follows:
$$L_{sur} = -\frac{1}{N_{E=1}} \sum_{i:E_i=1} (h_i - \log \sum_{j \in \mathcal{H}(T_i)} e^{h_j}), \quad (3)$$
where $h$ is the predicted risk of disease progression, $E$ is an event indicator (0 indicates a censored patient and 1 indicates a patient with disease progression), $T$ is the time of PFS (for $E = 1$) or the time of patient censored (for $E = 0$), $N_{E=1}$ is the number of patients with disease progression, and $\mathcal{H}(T_i)$ is a set of patients whose $T$ is no less than $T_i$.

## IV. EXPERIMENTAL SETUP

### A. Patients and Datasets

Our first dataset includes 170 patients acquired from Fudan University Shanghai Cancer Center (FUSCC). All patients were pathologically confirmed with advanced NPC (TNM stage III or IVa). The following endpoint is PFS or otherwise censored. All patients underwent [18F]fluorodeoxyglucose ([18F]FDG) PET/CT before treatment and have clinical indicators including age, gender, T stage, N stage, and TNM stage. We performed univariate Cox analyses for all clinical indicators and only the TNM stage showed significant relevance to PFS ($P = 0.047$). Therefore, only the TNM stage was used as a clinical feature in our experiments. More details can be found in our previous study [16] and in Section I of the Supplementary Materials.

Our second dataset includes 23 patients acquired from The Cancer Imaging Archive (TCIA). TCIA provides 298 patients with histologically confirmed Head-and-Neck Cancer [36]. All patients underwent [18F]FDG PET/CT before treatment. Details of the TCIA data can be found in [5]. 23 patients with advanced NPC were selected and were used to further validate the proposed method. The selection conditions are: (1) the primary site is nasopharynx and, (2) TNM stage is III or IVa.

The clinical characteristics of the patients enrolled in this study were shown in Table S1 of the Supplementary Materials.

### B. PET/CT Image Preprocessing

We preprocessed PET/CT images through resampling, SUV conversion (for PET), affine registration, Regions-of-Interest (ROIs) cropping, and intensity normalization (see Section II of the Supplementary Materials). Finally, we got PET/CT ROIs in 128×128×112 voxels, which cover the whole nasopharynx and are suitable for the segmentation and survival prediction of NPC. The PET/CT ROIs were used as the input of models, while the manually segmented masks of primary tumors were used as ground truth labels for the tumor segmentation task.

### C. Comparison Methods

Firstly, for comparison with radiomics-based survival models, we extracted 1456 radiomics features from PET/CT (see Section III of the Supplementary Materials). In addition, the clinical feature (TNM stage) was also used in all comparison methods as default if no extra setting is stated.

Then, our DeepMTS was compared to three feature-based survival models. Specifically, redundant radiomics features with Spearman's correlation > 0.7 were first eliminated for dimension reduction. Then, the remaining radiomics features and the clinical feature (TNM stage) were fed into two popular traditional survival models and one early deep survival model: (1) Lasso-Cox [38]: an extension of the CPH model [37] using LASSO regression for feature selection. (2) RSF (Random Survival Forest) [21]: a tree model that generates an ensemble estimate for survival data. (3) DeepSurv [20]: a fully-connected network using the same loss function as ours (Eq.3).

Moreover, our DeepMTS was also compared to three recent end-to-end deep survival models: (1) CNN-Survival [23]: a 2D CNN-based survival (CNN-Survival) model for survival prediction in pancreatic ductal adenocarcinoma. (2) MDSN [19]: a 3D deep survival network (MDSN), which is the most recent end-to-end deep model for survival prediction in NPC. (3) DLPM [24]: a deep learning survival prediction model (DLPM) for patients undergoing chest surgery for lung cancer. The MDSN and DLPM took the same input as our DeepMTS, while the CNN-Survival took as input 2D PET/CT slices that only contain manually segmented primary tumor regions. More details of these comparison methods can be found in Section IV of the Supplementary Materials.

Furthermore, we derived 236 deep features from the trained DeepMTS for comparison with handcrafted radiomics features. The deep features were derived from both the hard-sharing branch (124 features) and the CSN (112 features). Similarly, deep features with Spearman's correlation > 0.7 were first eliminated for dimension reduction. Then, the remaining deep features and the clinical feature (TNM stage) were fed into the Lasso-Cox, RFS, and DeepSurv.

### D. Experimental Settings

We performed a multi-task evaluation (in Section V-B) to explore how deep multi-task learning improves the DeepMTS using our hybrid multi-task architecture. In this experiment, we compared the DeepMTS to five degraded models in which one or more components were removed, and thus resembling multi-tasking learning in [29][31][34][35]: (1) Seg-Backbone: the segmentation backbone was used for tumor segmentation, while the CSN was removed; (2) Sur-HS: the hard-sharing branch was used for survival prediction, while the upsampling segmentation branch and the CSN were removed; (3) Sur-CasNet: the CSN was used for survival prediction, while the whole segmentation backbone was removed; (4) MT-HS: the segmentation backbone was used for both tumor segmentation and survival prediction, while the CSN was removed; (5) MT-CasNet: the segmentation backbone and CSN were used for both survival prediction and tumor segmentation, in which the hard-sharing branch was not used for survival prediction.

We explored different input strategies for the CSN to use tumor segmentation information (in Section V-C). In this experiment, two input strategies were explored in the Sur-CasNet and MT-CasNet: (1) Multiplication: tumor masks were voxel-wise multiplied with PET/CT to reveal primary tumors; (2) Concatenation: tumor masks were concatenated with PET/CT to offer additional information. The MT-CasNet used the tumor masks derived from the segmentation backbone during training, while the Sur-CasNet used manual tumor masks as the segmentation backbone has been removed.

We evaluated different segmentation backbones (in Section V-D) to explore how our segmentation backbone improves the DeepMTS. We established a baseline DeepMTS using the 3D U-net [32] as the segmentation backbone and compared the baseline DeepMTS to the original DeepMTS.

### E. Evaluation Metrics

The survival prediction task was evaluated using C-index [40] which measures the consistency between the predicted risk $h$ and the survival status $(E, T)$. A higher C-index indicates a more consistent prediction. For statistical analysis, a two-sided $P$ value < 0.05 is considered to indicate a statistically significant difference. The segmentation task was evaluated using Dice Similarity Coefficient (DSC) which measures the similarity between the predicted and ground-truth segmentation masks. A higher DSC generally indicates a better segmentation result.

### F. Implementation Details

Our DeepMTS was implemented using Keras with a Tensorflow backend on two 12GB Titan X GPUs. We used an Adam optimizer with a batch size of 8 to train the DeepMTS for 15000 iterations. The learning rate was 1e-4 initially and then decreased to 5e-5, 1e-5, and 1e-6 at the $2500^{th}$, $5000^{th}$, and $10000^{th}$ training iteration. All models were trained and validated through 5-fold cross-validation within the FUSCC set. During training, data augmentation was applied to the input images in real-time to avoid overfitting, including random translations up to 10 pixels, random rotations up to 5 degrees, and random flipping in the sagittal axis. Furthermore, since the majority of patients are right-censored (Table S1), we sampled an equal number of censored and uncensored samples during the data augmentation process to better leverage the information in uncensored samples. The 5 models trained and validated on 5-fold cross-validation were assembled (via normalization and average) and then tested on the TCIA set. Our implementation code is publicly available [1].

---

[1] https://github.com/MungoMeng/Survival-DeepMTS

## V. RESULTS

### A. Performance of Survival Prediction

The DeepMTS was compared with existing survival models for survival prediction, and the C-index results on the FUSCC and TCIA sets are shown in Table I. Grouping patients based on TNM stage (III or IVa) achieved a C-index of 0.563 on the FUSCC set but failed to derive a statistically significant result on the TCIA set ($P = 0.354$). All other survival models, leveraging both PET/CT and TNM stage, achieved significantly higher results than TNM stage grouping ($P < 0.05$). Among three survival models using radiomics features, the Radiomics features + Lasso-Cox achieved the highest results (C-index: 0.696/0.675), so it was regarded as a traditional benchmark. Among four end-to-end deep survival models, only DeepMTS achieved significantly higher results than the traditional benchmark (C-index: 0.722/0.698 vs 0.696 /0.675; $P = 0.016/0.027$), while the CNN-Survival, MDSN, and DLPM failed to outperform it (C-index: 0.658/0.628, 0.675 /0.639, and 0.667/0.642). The performance of DeepMTS was slightly degraded (C-index: 0.710/0.695) when TNM stage was not used. Furthermore, compared to radiomics features, deep features consistently resulted in significantly higher results in the Lasso-Cox, RFS, and DeepSurv (C-index: 0.711/0.692 vs 0.696/0.675, 0.707/0.688 vs 0.691/0.673, and 0.716/0.694 vs 0.687/0.666; $P = 0.026/0.045, 0.034/0.042$, and $0.012/0.027$).

### B. Multi-task Evaluation

The proposed DeepMTS was compared to five degraded models (Seg-Backbone, Sur-HS, Sur-CasNet, MT-HS, and MT-CasNet) for survival prediction and/or tumor segmentation, and the C-index/DSC results on the FUSCC set are shown in Table II. For the survival prediction task, our DeepMTS achieved the highest result (C-index: 0.722) and significantly outperformed the MT-HS and MT-CasNet (C-index: 0.706 and 0.698; $P = 0.042$ and $0.017$). The MT-HS and MT-CasNet also showed significantly higher results than their single-task counterparts, Sur-HS and Sur-CasNet (C-index: 0.678 and 0.674; $P = 0.008$ and $0.036$). For the tumor segmentation task, the MT-HS achieved the highest result (DSC: 0.763). Our DeepMTS gained the second-highest result, followed by Seg-Backbone and MT-CasNet (DSC: 0.760, 0.758, and 754).

We also visualized the feature maps of five degraded models and our DeepMTS in Fig. S2 of the Supplementary Materials. The weights of the feature map visualization indicate the contribution of corresponding voxels to the final predicted risk. The Sur-HS and Sur-CasNet cannot focus on primary tumors and derived much information from background tissues (Fig. S2c and Fig. S2d). Compared to the Sur-HS and Sur-CasNet, the MT-CasNet derived more information from primary tumors but still cannot accurately locate primary tumors (Fig. S2f). The MT-HS can derive information exactly from primary tumors but tended to overlook the prognostic information existing out of primary tumors (Fig. S2e). Our DeepMTS can accurately derive prognostic information from primary tumors while also capturing the prognostic information that exists out of primary tumors (Fig. S2g).

### C. Evaluation of Input Strategies

Two strategies of using tumor segmentation information (Multiplication and Concatenation) were explored in the Sur-CasNet and MT-CasNet for survival prediction, and the C-index results on the FUSCC set (5-fold cross-validation) are shown in Table III. The results of the Sur-CasNet taking as input only PET/CT (C-index: 0.674) or only manual tumor masks (C-index: 0.628) are also shown in Table III for

Table I. C-index results of comparison between the proposed DeepMTS and existing survival prediction models.

| Method | | FUSCC (5-fold cross-validation) | TCIA |
|---|---|---|---|
| TNM stage grouping | | 0.563 | / |
| Radiomics features | + Lasso-Cox | 0.696 | 0.675 |
| | + RFS | 0.691 | 0.673 |
| | + DeepSurv | 0.687 | 0.666 |
| Deep features | + Lasso-Cox | 0.711 | 0.692 |
| | + RFS | 0.707 | 0.688 |
| | + DeepSurv | 0.716 | 0.694 |
| CNN-Survival | | 0.658 | 0.628 |
| MDSN | | 0.675 | 0.639 |
| DLPM | | 0.667 | 0.642 |
| DeepMTS (no TNM stage) | | 0.710 | 0.695 |
| DeepMTS (Ours) | | **0.722** | **0.698** |

**Bold:** the highest result in each column is in bold.

Table II. Results of multi-task evaluation.

| Method | C-index | DSC |
|---|---|---|
| Seg-Backbone | / | 0.758 |
| Sur-HS | 0.678 | / |
| Sur-CasNet | 0.674 | / |
| MT-HS | 0.706 | **0.763** |
| MT-CasNet | 0.698 | 0.754 |
| DeepMTS | **0.722** | 0.760 |

**Bold:** the highest result in each column is in bold.

Table III. C-index results of comparison between two input strategies for the CSN to use tumor segmentation information.

| Method | Fold1 | Fold2 | Fold3 | Fold4 | Fold5 | Average |
|---|---|---|---|---|---|---|
| Sur-CasNet (only PET/CT) | 0.642 | 0.661 | 0.725 | 0.707 | 0.635 | 0.674 |
| Sur-CasNet (only Seg) | 0.594 | 0.614 | 0.693 | 0.660 | 0.579 | 0.628 |
| Sur-CasNet (Multiplication) | 0.611 | 0.646 | 0.712 | 0.684 | 0.607 | 0.652 |
| Sur-CasNet (Concatenation) | 0.596 | 0.680 | 0.693 | 0.731 | 0.582 | 0.656 |
| MT-CasNet (Multiplication) | 0.605 | 0.638 | 0.714 | 0.679 | 0.599 | 0.647 |
| MT-CasNet (Concatenation) | **0.658** | **0.688** | **0.765** | **0.737** | **0.642** | **0.698** |

**Bold:** the highest result in each column is in bold.

comparison. Compared to the Sur-CasNet using only PET/CT, the Multiplication strategy consistently resulted in significantly lower results in the Sur-CasNet and MT-CasNet (C-index: 0.652 and 0.647; $P$ = 0.016 and 0.009); the Concatenation strategy contributed to a significantly higher result in the MT-CasNet (C-index: 0.698; $P$ = 0.036) but a significantly lower result in the Sur-CasNet (C-index: 0.656; $P$ = 0.027).

### D. Evaluation of Segmentation Backbones

The baseline DeepMTS, using the U-net as the segmentation backbone, was compared to our original DeepMTS for both survival prediction and tumor segmentation, and the C-index/DSC results on FUSCC set are shown in Table IV. For the tumor segmentation task, the Seg-Backbone achieved a higher result than the U-net (DSC: 0.758 vs 0.744). Compared to the U-net backbone, our segmentation backbone consistently contributed to higher results in the MT-HS, MT-CasNet, and DeepMTS (DSC: 0.763 vs 0.747, 0.754 vs 740, and 0.760 vs 0.745). For the survival prediction task, compared with the U-net backbone, our segmentation backbone contributed to significantly higher results in the Sur-HS, MT-HS, and DeepMTS (C-index: 0.678 vs 0.655, 0.706 vs 0.680, and 0.722 vs 0.704; $P$ = 0.011, 0.021, and 0.039), while we did not identify significant difference between the original MT-CasNet and its U-net counterpart (C-index: 0.698 vs 0.696, $P$ > 0.05).

## VI. DISCUSSION

Our main findings are: (1) The DeepMTS outperformed all comparison methods on the survival prediction in advanced NPC and extracted more discriminative features than manually-defined feature extraction, (2) Deep multi-task learning contributed to better performance on survival prediction and our hybrid multi-task architecture outperformed existing hard-sharing and cascaded multi-task architectures, (3) Prognostic information exists both inside and outside of primary tumors and our DeepMTS can capture this information while getting less interference from non-relevant background information, and (4) Our segmentation backbone contributed to better performance on survival prediction and tumor segmentation.

In the comparison between our DeepMTS and existing survival models (Table I), our DeepMTS outperformed all comparison methods and showed a significantly higher C-index than the traditional benchmark (Radiomics features + Lasso-Cox). It should be noted that the traditional radiomics-based survival models required manual tumor segmentation for both training and testing, while the DeepMTS only requires it for training. We also found that, compared to the handcrafted radiomics features, the DeepMTS-derived deep features consistently contributed to a significantly higher C-index in the feature-based survival models (Table I). This demonstrates that the DeepMTS can extract more discriminative features than manually-defined feature extraction. We attribute this to the fact that the deep features derived from our DeepMTS can derive the prognostic information existing in the whole target regions (i.e., nasopharynx), while the radiomics features, extracted from segmented tumor regions, lost the prognostic information existing out of primary tumors. Moreover, we also identified that the end-to-end DeepMTS outperformed the feature-based survival models using deep features (Table I), which is consistent with the findings from Jing et al. [19] and Zhang et al. [23]. This is likely because end-to-end models allow cooperative learning of feature extraction and feature analysis, in which these two parts can better fit with each other.

Table IV. Results for evaluation of segmentation backbones.

| Method | C-index | DSC |
|---|---|---|
| U-net | / | 0.743 |
| Seg-Backbone (Ours) | / | **0.758 (+0.015)** |
| Sur-HS (U-net) | 0.655 | / |
| Sur-HS (Ours) | **0.678 (+0.023)** | / |
| MT-HS (U-net) | 0.680 | 0.747 |
| MT-HS (Ours) | **0.706 (+0.026)** | **0.763 (+0.016)** |
| MT-CasNet (U-net) | 0.696 | 0.740 |
| MT-CasNet (Ours) | **0.698 (+0.002)** | **0.754 (+0.014)** |
| DeepMTS (U-net) | 0.704 | 0.745 |
| DeepMTS (Ours) | **0.722 (+0.018)** | **0.760 (+0.015)** |

**Bold:** The results using our segmentation backbone are in bold with the improvements over their U-net counterparts in parentheses.

The CNN-Survival got the lowest C-index among all deep survival models (Table I). We attribute this to the fact that CNN-Survival is a 2D slice-level model that makes predictions based on a single slice of manually segmented primary tumor regions, which thereby disregards the potentially prognostic information that exists out of primary tumors and exists in 3D tumor volumes [10]. The MDSN and DLPM addressed the limitations of CNN-Survival by using a modified 3D DenseNet [17] which takes 3D images covering the whole target regions as the input. However, both MDSN and DLPM were inferior to the traditional benchmark in our experiments (Table I), which conflicts with the results reported by Jing et al. [19] and Kim et al. [24]. This is likely because the MDSN and DLPM were optimized for large datasets (1417 and 640 patients) [19][24] and thus tend to overfit to our relatively small dataset (170 patients). It should be noted that the overfitting issue was not identified in our DeepMTS, even though the DeepMTS is more complicated and has more learnable parameters than the MDSN and DLPM. We attribute this to the use of deep multi-task learning which improved the data efficiency by leveraging the shared features learned for multiple tasks [31]. This advantage is important in the medical domain because medical image data is usually scarce. PET/CT of NPC patients is even scarcer because MRI serves as a routine imaging tool for NPC patients, as it provides high soft-tissue resolution for assessing primary tumor [1], while PET/CT is acquired only when requiring molecular characterization (e.g., detection of distant metastasis).

In the multi-task evaluation (Table II), we identified that the MT-HS achieved a significantly higher C-index than the Sur-HS and a slightly higher DSC than the Seg-Backbone. These demonstrate that, in the hard-sharing multi-task architecture, deep multi-task learning enhanced the performance of survival prediction and did not impose negative impacts on the performance of tumor segmentation. This is because the tumor segmentation task can implicitly guide the hard-sharing branch to extract local features related to tumor regions, which allows the model to focus on primary tumors and get less interference from non-relevant background information. We also identified that the MT-CasNet achieved a significantly higher C-index than the Sur-CasNet and a slightly lower DSC than the Seg-Backbone (Table II). These demonstrate that, in the cascaded multi-task architecture, deep multi-task learning also enhanced the performance of survival prediction but slightly degraded the performance of tumor segmentation. This is likely attributed to the following two reasons: (1) The output of the segmentation backbone, providing global tumor information (e.g., tumor size, shape, and locations), was explicitly reused in the CSN to facilitate survival prediction; (2) The CSN makes segmentation target closer to the regions providing prognostic information, while the regions providing prognostic information are not essentially same as the manually segmented tumor regions (i.e., ground truth labels) [29]. For the same reason, the tumor segmentation performance of DeepMTS was degraded by the CSN and thus failed to outperform the MT-HS (Table II).

The multi-task evaluation (Table II) also showed that our DeepMTS significantly outperformed the MT-HS and MT-CasNet on survival prediction, which demonstrates that our hybrid multi-task architecture outperformed the commonly used hard-sharing and cascaded multi-task architectures. We attribute this to the fact that the hybrid multi-task architecture can benefit from the advantages of the two baseline multi-task architectures in the following two aspects: (1) The hybrid multi-task architecture can take advantage of tumor segmentation information both implicitly (through the hard-sharing backbone) and explicitly (through the CSN); (2) The hybrid multi-task architecture not only can keep its focus on primary tumor regions (through the hard-sharing backbone) but also can capture the prognostic information existing out of primary tumors (through the CSN).

In the evaluation of different input strategies for the CSN (Table III), multiplying PET/CT and tumor masks to reveal primary tumors (Multiplication strategy) consistently led to significantly lower results in the Sur-CasNet and MT-CasNet, which demonstrates that the prognostic information existing out of primary tumors has been lost. We also tried another input strategy, where we concatenated PET/CT and tumor masks (Concatenation strategy). Through this strategy, we expect models to get more complete information from PET/CT and also get global tumor information (e.g., tumor size, shape, and locations) from tumor masks. However, in the Sur-CasNet, the Concatenation strategy led to higher results only on the fold 2/4, while on the fold 1/3/5 it led to much lower results, where the low results are close to the ones derived from the Sur-CasNet using tumor masks only (Only Seg). We also observed that, on the fold 1/3/5, its training convergence became faster and approximates to the Sur-CasNet (Only Seg). Based on these observations, we suggest that the Concatenation strategy has the potential to improve models, but the Sur-CasNet using this strategy has the risk to overfit to manual tumor masks and ignore the information existing in PET/CT, as the binary tumor masks are much easier to fit than PET/CT, especially when our training set is small (136 patients for each fold). The MT-CasNet using the Concatenation strategy did not suffer from this problem and achieved significantly higher results. This is likely because the tumor masks used in the MT-CasNet are derived from the segmentation backbone, which are jointly trained and thus imposed inherent data augmentation on tumor masks. Based on these findings, we adopted the Concatenation strategy for the CSN in our DeepMTS.

In the evaluation of segmentation backbones (Table IV), our segmentation backbone consistently contributed to better performance on tumor segmentation than the U-net backbone. This demonstrates its superior capability in tumor segmentation. Moreover, compared to the U-net backbone, our segmentation backbone contributed to a significantly higher C-index in the Sur-HS and MT-HS (Table IV), indicating that the hard-sharing branch of our segmentation backbone has superior learning capability and can discover more discriminative features for survival prediction. However, our segmentation backbone did not cause a significant difference in the C-index of MT-CasNet ($P > 0.05$), which implies that the CSN is robust to the quality of the tumor masks derived from the segmentation backbone and thus is insensitive to its improvement. Nevertheless, the DeepMTS benefits from the improved hard-sharing branch and achieved better performance on survival prediction by using our segmentation backbone. We also evaluated different CSNs in Section V of the Supplementary Materials, in which we found that our CSN consistently contributed to a higher C-index than the CSN based on CNN-Survival [23]. This is expected as the DenseNet has been widely recognized in many image-related tasks including survival prediction [19][24].

There exist a few limitations in this study. The DeepMTS was trained and validated in a relatively small PET/CT dataset of 170 NPC patients (FUSCC set). To ensure reproducibility in real clinical environments, we have measured the statistical significance and tested the models in an external TCIA set. Nevertheless, a larger dataset is desired for further validation and potentially enables better performance. Moreover, we note that there are attempts at segmenting metastatic lymph nodes in NPC [41]. Our DeepMTS also can be trained to segment lymph nodes, and this could bring in potential benefits. Inclusion of lymph nodes into the segmentation target may enable the DeepMTS to capture the prognostic information existing in lymph nodes more precisely. Finally, our DeepMTS is not bounded to NPC and PET/CT. We suggest that the DeepMTS can be generalized to other cancers (e.g., head and neck cancer [42]) or other imaging modalities (e.g., MRI and CT).

## VII. CONCLUSION

In this study, we propose a deep multi-task framework, in which tumor segmentation, as an auxiliary task, is incorporated for survival prediction using a hybrid multi-task architecture. Under the proposed framework, we developed a 3D end-to-end Deep Multi-Task Survival model (DeepMTS) for survival prediction in NPC. The results with two clinical datasets have shown that the proposed deep multi-task framework improved the performance on survival prediction and our DeepMTS outperformed existing survival prediction models.